\documentclass[12pt,a4paper]{article}
\usepackage{hyperref}
\usepackage{lscape}
\usepackage{amsmath}
\usepackage{amssymb}
\usepackage{graphicx}
\usepackage{tensor}
\usepackage{romannum}
\usepackage{subcaption}
\usepackage{hyperref}
\usepackage{doi}
\usepackage{mathpazo}
\usepackage{graphicx}
\usepackage{amsmath,textcomp}
\usepackage{makeidx}
\makeindex

\usepackage{sectsty}
	\allsectionsfont{\sffamily\raggedright}
	\sectionfont{\sffamily\large\raggedright}
	\subsectionfont{\sffamily\normalsize\raggedright}
	
\usepackage{ftnright}

\usepackage{flushend}
\usepackage{cuted}
\usepackage{color}
\usepackage{graphicx}
\usepackage{hyperref}
\usepackage{pstricks,amssymb}
\usepackage{physics}
\usepackage{dsfont}
\usepackage{tensor}

\usepackage{fancyhdr}

\begin{document}
\title{\vspace{-2em}\bfseries\sffamily A sincere tribute to E.C.G Sudarshan's phenomenal contribution toward quantum theory of optical coherence }
\author{Arindam Kumar Chatterjee${^1}$, Anik Rudra${^2}$, Soham Chakraborty${^3}$\\[2ex]
	$^{1}$Department of Physics, West Bengal State University, \\Kolkata-700126, India.\\
	{\tt 72arindam@gmail.com}\\[2ex]
	$^{2}$Department of Physics, H. N. B. Garhwal University, \\S.R.T. Campus, Uttarakhand-249199, India.\\
	{\tt rudraanik13@gmail.com}\\[2ex]
	$^{3}$Department of Physics, University of Calcutta,\\ Kolkata-700073, India.\\
	{\tt sohamphy@yahoo.com}
}

\date{\today}

\maketitle

\thispagestyle{fancy}

\begin{abstract}
	{\sffamily
		The diagonal representation and optical equivalence theorem are the E. C. G. Sudarshan's mid 20th century adventures in non-classical optics. It basically deals with a quantum mechanical description of photons to explain the quantum properties of light. Inspired by Sudarshan's pioneering work we try to explain the every minute mathematical details of his paper "Equivalence of semi-classical and quantum mechanical descriptions of statistical light beams" published in [\ref{1}]. 
		
		In this article we are going to go through some of the basics in developing quantum optics, then land up in E.C.G's original work and try to present it as rigorous as possible. We show some of its important applications in various classes of physics problems.
	}

\end{abstract}

\begin{figure}[!h]
	\centering
	\includegraphics[height=5cm,width=7cm]{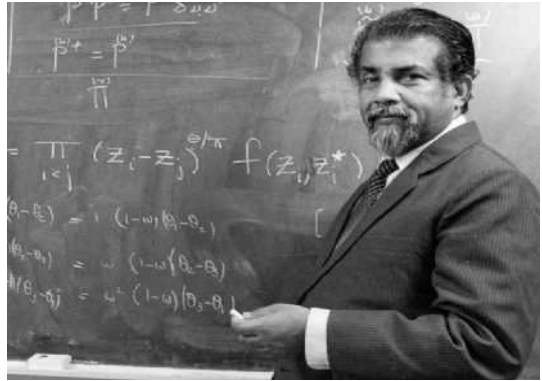}
	\caption{\small E. C. G. Sudarshan (16 September 1931 - 14 May 2018)}
\end{figure}

\newpage

\section{The fate of Irony in E.C.G's Physics Carrier}

Unlike the discovery of tachyon (a particle moves faster than light) and perhaps the most  popular, newsworthy issue of disproving  Einstein's proposal (nothing can move faster than light), many of  Sudarshan's path breaking works on various fields are much less popular even today in the science community. His research interest spanned over a wide range of field from Particle physics, Quantum optics, Quantum field theory to Quantum information theory, Gauge theory and Classical mechanics. In this paper we pay tribute to his contribution in particular to the quantum theory of optical coherence.

Sudarshan missed Nobel Prize several times. In 2005 Royal Swedish Academy Of Science chose to give Nobel prize to R. J. Glauber "for his contribution to the quantum theory of optical coherence". Although the credit for formulating and discovering the diagonal coherent state representation in its entirety and optical equivalence theorem must go to Sudarshan. It is truly ironic that these facts are readily accessible in [\ref{1}] and [\ref{2}], an expression which ought to be called "Sudarshan's diagonal coherent state representation" is dubbed as "Glauber's P-representation". Nor is it correct even one compromises himself/herself to call it the "Glauber-Sudarshan representation".

How long should an individual physicist be subjected to this kind of agony?. This is the point where only a large exchange of awareness required among the physics and non-physics community may give rise the solution. Sudarshan's work is not merely a mathematical formulation. It is the basic theory underlying all optical fields. All the quantum features are brought out automatically in this diagonal representation. However depth of his works can only be understood if one tries to solve his paper minutely. That is what we are going to present in the forthcoming section.

\section{Correspondence between linear harmonic oscillator and quantized electromagnetic field of radiation}

Why we discuss linear harmonic oscillator with such an emphasis? Just because there is an one to one correspondence between the linear harmonic oscillator problem and quantum optics i.e. the algebraic framework in which we discuss quantized electromagnetic field is the same as harmonic oscillator. In case of quantum optics we try to understand the various states of the quantized electromagnetic field of radiation as pointed below.

	Mapping between Simple Harmonic Oscillator and Quantified EM field.
	\begin{center}
		\begin{tabular}{|c|c|}
			\hline 
			Simple Harmonic Oscillator & Quantified EM field \\ 
			\hline 
			$\hat{a}$ : lowering operator & $\hat{a}$ : photon destruction operator \\ 
			\hline 
			$\hat{a}^{\dagger}$ : raising operator & $\hat{a}^{\dagger}$ : photon creation operator \\ 
			\hline 
			$\hat{n}$ : label of the energy eigenstate & $\hat{n}$ : no. of photons in a state \\
			\hline
		\end{tabular}  
	\end{center}

\section{Coherent state}

We need coherent state representation to interpret the state of a statistical beam of photons (i.e. electromagnetic wave). This state brings up a close relationship between the quantum and classical correlation function. Coherent state is an eigen state (with complex amplitude) of photon annihilation operator. Here $\hat{a}$ is not a Hermitian operator and its eigen value is some complex number. Then general coherent state is defined by
\begin{equation}
\hat{a}\ket{z}=z\ket{z}
\end{equation}
z is an general complex number. Then we must have
\begin{equation}
\bra{z}\hat{a}^{\dagger}=z^{*}\bra{z}
\end{equation}

\subsection{Fock state representation of coherent state}
Coherent state can be generated from vacuum state $\ket{0}$ by operating displacement operator on it. Let the displacement operator be
\begin{equation}
D(z)=e^{-\frac{|z|^2}{2}}e^{z\hat{a}^{\dagger}}e^{-z^{*}\hat{a}}
\end{equation}

The order is important because $\hat{a}$ and $\hat{a}^{\dagger}$ don't commute.

\begin{equation}
\begin{aligned}
\ket{z}&=e^{-\frac{|z|^2}{2}}e^{z\hat{a}^{\dagger}}e^{-z^{*}\hat{a}}\ket{0} \\
&=e^{-\frac{|z|^2}{2}}e^{z\hat{a}^{\dagger}}\ket{0}\\
&=e^{-\frac{|z|^2}{2}}\sum_{n=0}^{\infty}\frac{z^{n}}{\sqrt{n!}}\ket{n}\label{(3)}
\end{aligned}
\end{equation}

we have used the fact $e^{-z^{*}\hat{a}}\ket{0}=1\ket{0}$ and $\ket{n}=\frac{(\hat{a}^{\dagger})^{n}}{\sqrt{n!}}\ket{0}$, verificaton can be done by

\begin{equation*}
\begin{aligned}
\hat{a}(z)&=e^{-\frac{|z|^2}{2}}\sum_{n=0}^{\infty}\frac{z^{n}}{\sqrt{n!}}\hat{a}\ket{n} \\
&=e^{-\frac{|z|^2}{2}}\sum_{n=0}^{\infty}\frac{z^{n}}{\sqrt{n!}}\sqrt{n}\ket{n-1} \\
&=ze^{-\frac{|z|^2}{2}}\sum_{n=0}^{\infty}\frac{z^{n}}{\sqrt{n!}}\ket{n} \\
&=z\ket{z} \\
&\text{$\because\hat{a}\ket{n}=\sqrt{n}\ket{n-1}$}
\end{aligned}
\end{equation*}

\section{Coherent state representation of number state}

If we multiply eq. (\ref{(3)}) by $\frac{1}{\pi}\frac{1}{\sqrt{m!}}z^{*m}e^{-\frac{|z|^2}{2}}$ and integrating both side over z we get 
\begin{equation*}
\begin{aligned}
&\frac{1}{\pi}\int\frac{1}{\sqrt{m!}}z^{*m}e^{-\frac{|z|^2}{2}}\ket{z}d^2z \\
=&\frac{1}{\pi}\sum_{n=0}^{\infty}\int e^{-|z|^2}\frac{z^{n}}{\sqrt{n!}}\frac{z^{*m}}{\sqrt{m!}}\ket{n}d^2z
\end{aligned}
\end{equation*}

Here we use polar co-ordinate $z=re^{i\theta}$ and $d^2z=r\:dr\:d\theta$ then we directly perform $\theta$ integral using the identity

\begin{equation*}
\int_{\theta=0}^{2\pi}e^{i(n-m)\theta}d\theta=2\pi\delta_{nm}
\end{equation*}

and the rest of it is straight forward $\Gamma$ function integral. We obtain the following

\begin{equation}
\begin{aligned}
&\frac{1}{n!}\Gamma(n+1)\ket{n}=\ket{n} \\
&\ket{n}=\frac{1}{\pi}\int\frac{1}{\sqrt{n!}}z^{*n}e^{-\frac{|z|^2}{2}}\ket{z}d^{2}z
\end{aligned}
\end{equation}

Again we have used $\displaystyle\int_{y=0}^{\infty}e^{-y^m}y^{n}dy=\frac{1}{m}\Gamma\left(\frac{n+1}{n}\right)$ and $\Gamma\left(n+1\right)=n!$
From eq. (\ref{(3)}) it is clear that projection of $\ket{z}$ on every Fock state gives a non zero value for all complex number (other than zero). Thus

\begin{equation}
\bra{n}\ket{z}=e^{-\frac{|z|^2}{2}}\frac{z^{n}}{n!}
\end{equation}

and probability that n excitation or photons will be found in the coherent state $\ket{z}$ can be expressed as 

\begin{equation*}
P(n)=|\bra{n}\ket{z}|^2=e^{-|z|^2}\frac{z^{2n}}{n!}
\end{equation*}

The average number of photons for the state $\ket{z}$ is found to be

\begin{equation*}
\bar{n}=\sum_{n=0}^{\infty}nP(n)=\bra{z}\hat{a}^{\dagger}\hat{a}\ket{z}=|z|^2
\end{equation*}

This average number of photons depends on $|z|$. So, the average number of photon is large or small as the $|z|$ value is either large or small. But no matter how small $|z|$ may be (except $z=0$) there is always some nonzero probability $P(n)$, which means any number of photon $'n'$ is present in field.
If we express $P(n)$ in terms of $\bar{n}$ then, $P(n)=e^{-\bar{n}}\frac{\bar{n}^n}{n!}$ which is a Poisson distribution. Where the variance in photon number is $(\Delta N)^2$.

Here the remarkable feature of coherent state is that the state remains unchanged when annihilation operator acting on it. It predicts a connection between quantum and classical fields. It suggests that it is possible to absorb photons from electromagnetic
field in coherent state repeatedly without changing the state any way.

\section{Properties of coherent states}

The coherent state carries some interesting features, We will show some of them.

\begin{description}
	\item[$\bullet$] They satisfy Heisenberg’s minimum uncertainty product.
	\item[$\bullet$] They follow closure property.
	\item[$\bullet$] They form an over-complete basis set but they are not mutually orthogonal to each other.
\end{description}

\subsection{Heisenberg’s minimum uncertainty relation of coherent state}
This property is very trivial, We will leave it without proving. We will focus on the other two properties, in fact properties (ii) and (iii) leads us that coherent states can be used as a basis set.

\subsection{Closure Property of coherent state}

Though coherent states are not orthogonal, but they follow closure property. It is as
follows
\begin{equation*}
\begin{aligned}
		&\int d^2z\ket{z}\bra{z} \\
		&=\int d^2z\: e^{-|z|^2}\sum_{n=0}^{\infty}\sum_{m=0}^{\infty}\frac{z^{n}}{\sqrt{n!}}\frac{z^{*m}}{\sqrt{m!}}\ket{n}\bra{m}
\end{aligned}
\end{equation*}

using polar coordinate $z=re^{i\theta}$ and performing $\theta$ integral we get

\begin{equation*}
\begin{aligned}
&=\sum_{n=0}^{\infty}\frac{\ket{n}\bra{n}}{n!}\:2\pi\int_{r=0}^{\infty}e^{-r^2}r^{2n+1}dr \\
&=\sum_{n=0}^{\infty}\frac{\ket{n}\bra{n}}{n!}\:2\pi\:\frac{1}{2}\Gamma\left(n+1\right) \\
&=\pi\sum_{n=0}^{\infty}\ket{n}\bra{n} = \pi\times\hat{\mathds{1}} \\
&\text{So, $\frac{1}{\pi}d^2z\ket{z}\bra{z}=\hat{\mathds{1}}$}
\end{aligned}
\end{equation*}

\subsection{Non-orthogonality relations of coherent state}
To verify whether the coherent states are orthogonal or not, we take two different
coherent states $\ket{z_{1}}$ and $\ket{z_{2}}$, taking innerproduct of two states we get

\begin{equation}
\begin{aligned}
\bra{z_{1}}\ket{z_{2}}&=e^{-\frac{|z_{1}|^2}{2}}e^{-\frac{|z_{2}|^2}{2}}\sum_{n=0}^{\infty}\sum_{m=0}^{\infty}\frac{z_{1}^{*n}}{\sqrt{n!}}\frac{z_{2}^{m}}{\sqrt{m!}}\bra{n}\ket{m} \\
&=e^{-\frac{|z_{1}|^2}{2}}e^{-\frac{|z_{2}|^2}{2}}\sum_{n=0}^{\infty}\frac{z_{1}^{*n}z_{2}^{n}}{n!} \\
&=e^{\frac{-1}{2}\left(|z_{1}|^2+|z_{2}|^2\right)}e^{z_{1}^{*}z_{2}} \\
&\neq{0}
\end{aligned}
\end{equation}

we have used the fact $\bra{n}\ket{m}=\delta_{nm}$ and $\displaystyle\sum_{n=0}\frac{x^{n}}{n!}=e^{x}$. 

Therefore the projection probability of $\ket{z_{1}}$ on $\ket{z_{1}}$ is

\begin{equation}
	\begin{aligned}
		|\bra{z_{1}}\ket{z_{2}}|^2&=\bra{z_{1}}\ket{z_{2}}\bra{z_{2}}\ket{z_{1}} \\
		&=e^{-|z_{1}|^2-|z_{2}|^2+z_{1}^{*}z_{2}+z_{1}z_{2}^{*}} \\
		&=e^{-|z_{1}-z_{2}|^2}
	\end{aligned}
\end{equation}

This indicated that, if a system is in state $\ket{z_{1}}$ then there is a nonzero probability that the system will be found in any other quantum state $\ket{z_{2}}$. In other words it is also said that coherent states are linearly dependent. This is so because one coherent state can be represented in terms of the other states. Let us take a coherent state $\ket{z^{\prime}}$
\begin{equation}
	\begin{aligned}
		\ket{z^{\prime}}&=\frac{1}{\pi}\int\ket{z}\bra{z}\ket{z^{\prime}}\:d^2z \\
		&=\frac{1}{\pi}\int\ket{z}e^{\frac{-|z-z^{\prime}|^2}{2}}e^{\frac{z^{*}z^{\prime}-z\:z^{*\prime}}{2}}\:d^2z
	\end{aligned}
\end{equation}

Therefore any one coherent state can be represented by all of them. Actually they form an over-complete set.

\section{What does over-completeness mean }

Properties of over-completeness can easily be understood if we take an example. Let’s consider a vector $\vec{r}$ which is a position vector of a point in $\mathbb{R}^2$ vector space. $\hat{e_{1}}$, $\hat{e_{2}}$ are the set of orthonormal vectors. We can represent $\vec{r}$ as
\begin{equation*}
	\vec{r}=c_{1}\hat{e_{1}}+c_{2}\hat{e_{2}}
\end{equation*}
This set of components $c_{1}$, $c_{2}$ are unique. If we consider two non-orthogonal unit vectors $\hat{e_{1}}^{\prime}$ and $\hat{e_{2}}^{\prime}$ as basis then the same position vector $\vec{r}$ can be represented as

\begin{equation*}
	\vec{r}={c_{1}}^{\prime}\hat{e}^{\prime}_{1}+c_{2}^{\prime}\hat{e}^{\prime}_{2}
\end{equation*}

but it is also unique. And if we consider three non-orthogonal unit vectors $\hat{e}^{\prime\prime}_{1}$, $\hat{e}^{\prime\prime}_{2}$, $\hat{e}^{\prime\prime}_{3}$ as basis although they are dependent then we can represent

\begin{equation*}
		\vec{r}=c_{1}^{\prime\prime}\hat{e}^{\prime\prime}_{1}+c_{2}^{\prime\prime}\hat{e}^{\prime\prime}_{2}+c_{2}^{\prime\prime}\hat{e}^{\prime\prime}_{3}
\end{equation*}

The set of components $c_{1}^{\prime\prime}$, $c_{2}^{\prime\prime}$, $c_{3}^{\prime\prime}$ are not unique. Uniqueness is lost here. Now we conclude the fact that if in a vector space of more number of basis vectors although they are dependent and also non-orthogonal are given then we can’t represent a vector not in a unique way. That basis set is not only complete but it is over-complete. Here in this example the basis is finite and the over-completeness can be removed by the removal of one of base vector. But it is not so easy to deal with the infinite no of basis formed by the coherent states.

Because of non-othogonality property of coherent state, we have to face some problem to interpret the projection probability of coherent state $\ket{z}$ and some other state $\ket{\phi}$. First we try with a Fock state $\ket{n}$. Here the scalar product $\bra{n}\ket{\phi}$ represents the probability amplitude of $\ket{\phi}$ in n-representation, which gives $|\bra{n}\ket{\phi}|^2$, the probability of finding n photons. Now, the Fock states are orthogonal and the probabilities $|\bra{n}\ket{\phi}|^2$ are mutually exclusive for different n and this gives

\begin{equation*}
	\sum_{n=0}^{\infty}|\bra{n}\ket{\phi}|^2=\sum_{n=0}^{\infty}\bra{\phi}\ket{n}\bra{n}\ket{\phi}=\bra{\phi}\ket{\phi}=1
\end{equation*}

On the other hand for coherent state the square of the scalar products $|\bra{z}\ket{\phi}|^2$ do not represent mutually exclusive probabilities (rather probability densities) and do not integrate to unity.

\begin{equation*}
	\int|\bra{z}\ket{\phi}|^{2} d^2z=\int\bra{\phi}\ket{z}\bra{z}\ket{\phi}d^2z = \pi
\end{equation*}

In duducing the above equation we have used the fact $\frac{1}{\pi}\int d^2z \ket{z}\bra{z}=1$. Here, if $|\bra{z}\ket{\phi}|^{2}$ represents the probability density of finding the complex amplitude z, then differential probabilities are evidently not mutually exclusive and the right hand side of the above equation gives an idea of the degree of overlap. 

For coherent states, non-orthogonality is significant mainly for neighbouring states and two states having different eigen values are almost orthogonal. 
Thus, the non-exclusiveness of the probability densities $|\bra{z}\ket{\phi}|^{2}$ is associated mainly with neighbouring coherent states and if $\ket{z_{1}}$ and $\ket{z_{2}}$ are two different states with appreciably different eigen values $z_{1}$, $z_{2}$ of $\hat{a}$ but $\frac{|\bra{z_{1}}\ket{\phi}|^{2}d^2z}{\pi}$ and $\frac{|\bra{z_{2}}\ket{\phi}|^{2}d^2z}{\pi}$ play the role of almost mutually exclusive differential probabilities. 

\section{More on over-complete basis set}

Consider the matrix elements of an operator $\hat{A}$ in Fock basis. The matrix elements are $\bra{n}\hat{A}\ket{m}$, where $m, n=0,1,2,3\cdots\infty$ exhaust all matrix elements. It is seen that all the diagonal and non-diagonal matrix elements expressed in Fock basis can be written down in coherent state basis using only diagonal terms i.e. $\bra{z}\hat{A}\ket{z}$ as

	\begin{equation}
		\bra{n}\hat{A}\ket{m}=\frac{1}{\sqrt{n!}\sqrt{m!}}\left(\frac{\partial^{n+m}}{\partial z^{*n}\partial z^{m}}\bra{z}\hat{A}\ket{z}e^{|z|^{2}}\right)\bigg|_{z^{*},z=0}
	\end{equation}

This suggests that only diagonal matrix elements in coherent state basis are enough to write all the diagonal and non-diagonal matrix elements written in Fock basis. This is the essence of over completeness property of coherent state basis. It is actually more than complete. Let us take an example for $n=2$ and $m=3$. So the corresponding matrix element is $\bra{2}\hat{A}\ket{3}$. Now from the above equation we have

	\begin{equation}
		\begin{aligned}
			\bra{n}\hat{A}\ket{m}&=\frac{1}{\sqrt{n!}\sqrt{m!}}\left(\frac{\partial^{n+m}}{\partial z^{*n}\partial z^{m}}\bra{z}\hat{A}\ket{z}e^{|z|^{2}}\right)_{z^{*},z=0} \\
			&=\frac{1}{\sqrt{n!}\sqrt{m!}}\left(\frac{\partial^{n+m}}{\partial z^{*n}\partial z^{m}} \ e^{-|z|^{2}}\sum_{p=0}^{\infty}\sum_{q=0}^{\infty}\frac{z^{*p} \ z^{q}}{\sqrt{p!}\sqrt{q!}}\bra{p}\hat{A}\ket{q}e^{|z|^{2}}\right)_{z^{*},z=0} \\
			&=\frac{1}{\sqrt{n!}\sqrt{m!}}\left(\frac{\partial^{n+m}}{\partial z^{*n}\partial z^{m}}\sum_{p=0}^{\infty}\sum_{q=0}^{\infty}\frac{z^{*p} \ z^{q}}{\sqrt{p!}\sqrt{q!}}\bra{p}\hat{A}\ket{q}\right)_{z^{*},z=0} \\
			\text{Putting $n=2$ and $m=3$} \\
			&=\frac{1}{\sqrt{2!}\sqrt{3!}}\left(\frac{\partial^{2+3}}{\partial z^{*2}\partial z^{3}}\sum_{p=0}^{\infty}\sum_{q=0}^{\infty}\frac{z^{*p} \ z^{q}}{\sqrt{p!}\sqrt{q!}}\bra{p}\hat{A}\ket{q}\right)_{z^{*},z=0}
		\end{aligned}
	\end{equation}

All the terms come from the summation do not contribute to get the matrix element. Only the term with $p=2$ and $q=3$ contributes here. This is so because the terms with $p<2$ and $q<3$ reduces to zero as the order of differentiation in each is higher than they appear. On the other hand the terms with $p>2$ and $q>3$ vanishes because of the restriction $z^{*},z=0$. Thus we get

	\begin{equation}
		\begin{aligned}
			\bra{2}\hat{A}\ket{3}&=\frac{1}{\sqrt{2!}\sqrt{3!}}\left(\frac{\partial^{2+3}}{\partial z^{*2}\partial z^{3}}\frac{z^{*2} \ z^{3}}{\sqrt{2!}\sqrt{3!}}\bra{2}\hat{A}\ket{3}\right)_{z^{*},z=0} \\
			&=\frac{1}{2!\cdot3!}\left(\frac{\partial^{2+3}}{\partial z^{*2}\partial z^{3}} \ z^{*2} \ z^{3}\right)_{z^{*},z=0}\bra{2}\hat{A}\ket{3} \\
			&=\frac{1}{2!\cdot3!}\times2!\cdot3!\times\bra{2}\hat{A}\ket{3} \\
			&=\matrixel{2}{\hat{A}}{3}
		\end{aligned}
	\end{equation}

\section{A comprehensive review of Sudarshan's paper}

In this section we give an elaborate calculation of E.C.G. Sudarshan's one of the famous paper on quantum optics. It was published in 1963 and entitled as "Equivalence of semi-classical and quantum descriptions of statistical beams". According to this paper- classical theory of optical coherence" is adequate for the description of the classical optical phenomena of interference and diffraction in general. More sophisticated experiments on intensity interferometry and photo electric counting statistics necessitated special higher order correlations. Most of this work was done using classical or semi classical formulation of the problems. On the other hand, statistical states of quantized field (electromagnetic) have been considered recently, and a quantum mechanical definition of coherence function of arbitrary order represented." The aim of this paper is to elaborate quantum definition of coherence function and find complete equivalence to classical description.

We have the ladder operators $\hat{a}$ and $\hat{a}^{\dagger}$, they obey the commutation relation $\left[\hat{a},\hat{a}^{\dagger}\right]=\hat{\mathds{1}}$ and number operator $\hat{N}$ is defined by $\hat{N}=\hat{a}^{\dagger}\hat{a}$. State kets corresponding to number operator i.e. Fock state obey the relation,
\begin{equation}
	\begin{aligned}
			\hat{a}^{\dagger}\hat{a}\ket{n}=\hat{N}\ket{n}&=n\ket{n} \\
			\bra{n}\ket{m}&=\delta_{n,m}
	\end{aligned}
\end{equation}
Matrix elements of $\hat{a}$ and $\hat{a^{\dagger}}$ are 
\begin{equation}
		\bra{n}\hat{a^{\dagger}}\ket{m}=\sqrt{m}\delta_{n,m-1}
\end{equation}
and
\begin{equation}
	\bra{n}\hat{a^{\dagger}}\ket{m}=\sqrt{m+1}\delta_{n,m+1}
\end{equation}
And the coherent state $\ket{z}$ in Fock state repretation is 
\begin{equation}
		\ket{z}=e^{-\frac{|z|^2}{2}}\sum_{n=0}^{\infty}\frac{z^{n}}{\sqrt{n!}}\hat{a}\ket{n},\medskip\text{where} \ n =0,1,2,\cdots\label{12}
\end{equation}
We know that $z$ is a complex number and it is $z=re^{i\theta}$. This coherent states are normalized but not orthogonal, satisfy closure relation and form an over complete set. Now we use this over completeness property to find the density matrix which is defined as
\begin{equation}
		\hat{\rho}=\sum_{n=0}^{\infty}\sum_{m=0}^{\infty}\rho\left(n,m\right)\ket{n}\bra{m}
\end{equation}
It is not diagonal representation. Now using coherent states, we will get a diagonal representation which can describe the whole density matrix. Now multiplying both sides of eq. (\ref{12}) by $e^{\frac{z^2}{2}}$ and taking outer product with $\bra{z}$ we get
\begin{equation*}
	e^{-\frac{|z|^2}{2}}\ket{z}\bra{z}=\sum_{n,m=0}^{\infty}\frac{z^{n}}{\sqrt{n!}}\frac{z^{*m}}{\sqrt{m!}}\ket{n}\bra{m}
\end{equation*}
using polar co-odinate $z=re^{i\theta}$ we have
\begin{equation*}
\begin{aligned}
		&e^{r^{2}}\ket{re^{i\theta}}\bra{re^{i\theta}} \\
		=&\sum_{n,m=0}^{\infty}\frac{r^{n+m}}{\sqrt{n!}\sqrt{m!}} \ e^{i\left(n-m\right)\theta}\ket{n}\bra{m}
\end{aligned}
\end{equation*}
Now multiplying both sides by $e^{i\theta}$ and integrating over $\theta$ in the range $\theta=0\:to\:2\pi$
\begin{equation*}
	\begin{aligned}
		&\int_{\theta=0}^{2\pi}\frac{d\theta}{2\pi}e^{r^{2}}e^{il\theta}\ket{re^{i\theta}}\bra{re^{i\theta}} \\
		&=\sum_{n,m=0}^{\infty}\frac{r^{n+m}}{\sqrt{n!}\sqrt{m!}}\delta\left(l+n-m\right)\ket{n}\bra{m}
	\end{aligned}
\end{equation*}
Now operate $\frac{\partial}{\partial r}$ for $p$ times on both sides and evaluate it  at $r=0$. It can be checked easily that the R.H.S of the new expression would be of physical if $n+m\geq p$ from one delta function. The other delta function picks at $l+n=m$. Therefore we get, \\
\\

	\begin{equation}
		\ket{n}\bra{m}=\frac{\sqrt{n!}\sqrt{m!}}{\left(n+m\right)!}\left(\frac{\partial}{\partial r}\right)^{m+n}\int_{\theta=0}^{2\pi}\frac{d\theta}{2\pi}\:e^{r^{2}}e^{i\left(n-m\right)\theta}\ket{re^{i\theta}}\bra{re^{i\theta}}\Bigg|_{r=0} \label{14}
	\end{equation}

The density operator in number state representation
\begin{equation}
	\begin{aligned}
	&\hat{\rho}=\sum_{n,m=0}^{\infty}\rho_{n,m}\ket{n}\bra{m} \\
	\text{where,} \\
	&\rho_{n,m}=\bra{n}\hat{\rho}\ket{m} \label{15}
	\end{aligned}
\end{equation}

This equation infers that any matrix written down in Fock basis can be represented also by coherent state basis. But unique feature is that in the coherent state basis we can only need diagonal matrix elements.

Let's see how this is possible. Taking trace to both side of eq. (\ref{14}) we get

	\begin{equation}
		\bra{n}\hat{O}\ket{m}=\frac{\sqrt{n!}\sqrt{m!}}{\left(n+m\right)!}\left(\frac{\partial}{\partial r}\right)^{m+n}\int_{\theta=0}^{2\pi}\frac{d\theta}{2\pi} \ e^{r^{2}}e^{i\left(n-m\right)\theta}\bra{re^{i\theta}}\hat{O}\ket{re^{i\theta}}\Bigg|_{r=0}	
	\end{equation}

From the above equation we see that diagonal or non-diagonal matrix element of expressed in Fock basis can be represented by considering only the diagonal matrix element in coherent state basis. Now substituting eq. (\ref{14}) in eq. (\ref{15}) we get

	\begin{equation}
		\begin{aligned}
			\hat{\rho}=\sum_{n,m=0}^{\infty}\rho_{n,m}\frac{\sqrt{n!}\sqrt{m!}}{\left(n+m\right)!}\int_{\theta=0}^{2\pi}\frac{d\theta}{2\pi} \ e^{i\left(n-m\right)\theta}\frac{\partial^{^{m+n}}}{\partial r^{^{m+n}}}e^{r^2}\ket{z}\bra{z}\Bigg|_{r=0} \\
			\text{using the property$\colon\int_{x=-\infty}^{+\infty}dx \ \delta^{(n)}(x)\varphi(x)=(-1)^{n}\varphi^{(n)}(0)$ we get}\label{19}
		\end{aligned}
	\end{equation}

	\begin{equation}
	\hat{\rho}=\int_{r=0}^{\infty}r \ dr\int_{\theta=0}^{2\pi}d\theta\ket{z}\bra{z}\frac{e^{r^2}}{2\pi r}\sum_{n,m=0}^{\infty}\rho_{n,m}\frac{\sqrt{n!}\sqrt{m!}}{\left(n+m\right)!}(-1)^{n+m} \ e^{i\left(n-m\right)\theta} \ \frac{\partial^{^{m+n}}}{\partial r^{^{m+n}}} \ \delta(r)\label{20}
	\end{equation}

This is the required diagonal representation. It has a greater significance when we take an expectation of a normal ordered operator i.e. $\hat{A}=a^{\dagger p}a^{q}$ in the statistical state represented by the density operator $\hat{\rho}$.

\begin{equation}
	\hat{\rho}=\frac{1}{\pi}\int d^2z\ket{z}\bra{z}\varphi(z)
\end{equation}

The statistical state given above are written in diagonal coherent state basis due to the property of over-completeness.
The expectation value of $\langle\hat{A}\rangle$

\begin{equation}
	\begin{aligned}
		\langle\hat{A}\rangle&=tr\left[\hat{\rho}\hat{A}\right] \\
		&=\frac{1}{\pi}\int d^2z \ \varphi(z)\matrixel{z}{\hat{A}}{z} \\
		&=\frac{1}{\pi}\int d^2z \ \varphi(z) \ z^{*p} \ z^{q}
	\end{aligned}
\end{equation}
we have used $a\ket{z}$ = $z\ket{z}$ and $a^{q}\ket{z}$ = $z^{q}\ket{z}$.

It represents the expectation value of a complex classical function $f\left(z^{*},z\right)$ = $z^{*p} \ z^{q}$ for a distribution function $\varphi(z)$ over the complex plane. So any statistical state of quantum mechanical system may be described by a classical probability distribution over a complex plane. Thus to get the expectation value we have to find out the classical distribution. However $\varphi(z)$ cannot be so easily interpreted, For classical state of light is a probability density. Classical probability density is non-negative and cannot be more singular than delta function. For non-classical state $\varphi(z)$ is not well behaved as classical probability density  and $\varphi(z)$ may be negative and it is easy to see if we take an example of Wigner distribution (1932). But it is beyond the purpose of this paper. Now it is easy to find $\varphi(z)$, we have

\begin{equation}
	\hat{\rho}=\int d^2z\ket{z}\bra{z}\varphi(z)
\end{equation}

comparing above expression to eq.(\ref{20}) we get

	\begin{equation}
		\begin{aligned}
			\varphi(z)&=\frac{e^{r^2}}{2\pi r}\sum_{n,m=0}^{\infty}\rho_{n,m}\frac{\sqrt{n!}\sqrt{m!}}{\left(n+m\right)!}(-1)^{n+m} \ e^{i\left(n-m\right)\theta} \ \frac{\partial^{^{m+n}}}{\partial r^{^{m+n}}} \ \delta(r) \\
			&=\frac{e^{r^2}}{2\pi r}\left[\sum_{n,m=0}^{\infty}\rho_{n,m} \ \frac{n!}{(2n)!} \ \frac{\partial^{^{2n}}}{\partial r^{^{2n}}} \ \delta(r)+\left(\text{terms contaning $n\neq m$}\right)\right]\label{24}
		\end{aligned}
	\end{equation}

which is required distribution function.

\section{Normalization of $\varphi(z)$}

\begin{equation}
	\int d^2z \ \varphi(z)=\int_{r=0}^{\infty}r \ dr\int_{\theta=0}^{2\pi}\ \varphi(z) \ d\theta
\end{equation}

from eq.(\ref{24}) it is clear that $\varphi(z)$ integration exist only when $n=m$. Thus

\begin{equation}
	\begin{aligned}
		\int d^2z \ &\varphi(z) \\
		&=\sum_{n=0}^{\infty}\rho_{n,n} \ \frac{n!}{(2n)!}\int_{r=0}^{\infty}dr \ e^{r^2}\frac{\partial^{^{2n}}}{\partial r^{^{2n}}}\delta(r) \\
		&=\sum_{n=0}^{\infty}\rho_{n,n}=1
	\end{aligned}
\end{equation}

Here we have used property mention in eq. (\ref{19}). It says that this distribution function is normalized.

\section{In search of $\varphi(z)$ for three different classes of physics problem}

\subsection{Problem 1$\colon$}

Given $\hat{\rho}=\ket{0}\bra{0}$, We can find $\varphi(z)$. At first we have to find the matrix element $\hat{\rho}_{n,m}=\matrixel{n}{\hat{\rho}}{m}$
\begin{equation}
	\begin{aligned}
		\hat{\rho}_{n,m}&=\matrixel{n}{\hat{\rho}}{m}=\braket{n}{0}\braket{0}{m} \\
		&=1 ,\ \text{for $n=m=0$} \\
		&=0 ,\ \text{otherwise}
	\end{aligned}
\end{equation}

Therefore, putting $n=m=0$ in eq. (\ref{24}) we can get the possible distribution function as

\begin{equation}
	\varphi(z)=\frac{e^{r^2}}{2\pi r}\delta(r)=-\frac{e^{r^2}}{2\pi} \ \delta^{1}(r)
\end{equation}
Where $\delta^{1}(r)$ is the first derivative of $\delta(r)$.

\subsection{Problem 2$\colon$}

Given $\hat{\rho}=\ket{z}\bra{z}$, we can find $\varphi(z)$.
\begin{equation}
	\begin{aligned}
		\hat{\rho}_{n,m}&=\matrixel{n}{\hat{\rho}}{m} \\
		&=\braket{n}{z}\braket{z}{m} \\
		&=\sum_{n,m=0}^{\infty}e^{-|z|^{2}}\frac{z^{n}}{\sqrt{n!}}\frac{z^{*m}}{\sqrt{m!}}
	\end{aligned}
\end{equation}

since we know $z=re^{i\theta}$ and $|z|=r$ in polar co-ordinate. Now substituting $\rho_{n,m}$ in eq. (\ref{24}) we get

	\begin{equation}
		\varphi(z)=\frac{1}{2\pi}\sum_{n,m=0}^{\infty}(-1)^{n+m} \ \frac{r^{n+m-1} \ e^{2i\left(n-m\right)\theta}}{\left(n+m\right)!}\frac{\partial^{^{m+n}}}{\partial r^{^{m+n}}} \ \delta(r)
	\end{equation}

which is our required distribution function.

\subsection{Problem 3$\colon$}

Now we will discuss the phase space distribution $\varphi(z)=\frac{1}{\pi\bar{n}}e^{\frac{-|z|^2}{\bar{n}}}$ corresponds to the density operator $\hat{\rho}$ for radiation field in thermal equilibrium at temperature T. Following the same procedure as that of our previous two cases and using polar co-ordinate we get

	\begin{equation}
		\begin{aligned}
			\hat{\rho}&=\int d^2z\ket{z}\bra{z}\varphi(z) \\
			&=\sum_{n=0}^{\infty}\sum_{m=0}^{\infty}\int \frac{1}{\pi\bar{n}} \ e^{\frac{r^2}{\bar{n}}} \ \frac{r^{n+m} \ e^{i\left(n-m\right)\theta}}{\sqrt{n!}\sqrt{m!}}e^{-r^2}\ket{n}\bra{m}r \ dr \ d\theta \\
		\end{aligned}
	\end{equation}

separating $r$, $\theta$ integration and using delta function identity we get,

\begin{equation*}
	\begin{aligned}
		\hat{\rho}&=2\sum_{n=0}^{\infty}\frac{\ket{n}\bra{n}}{\bar{n}\:n!}\int_{0}^{\infty}e^{-r^2}\left(\frac{1}{\bar{n}}+1\right)r^{2n+1}dr \\
		&=\sum_{n=0}^{\infty}\frac{\ket{n}\bra{n}}{\bar{n}\:n!}\times\frac{\Gamma\left(n+1\right)}{\left(1+\frac{1}{\bar{n}}\right)^{n+1}} \\
		&=\sum_{n=0}^{\infty}\frac{1}{\left(\bar{n}+1\right)}\left(1+\frac{1}{\bar{n}}\right)^{-n}\ket{n}\bra{n}
	\end{aligned}
\end{equation*}

we have used $\displaystyle\int_{0}^{\infty}e^{-ax^{m}}x^{n}dx$ = $\frac{1}{m}\frac{\Gamma\left(\frac{n+1}{m}\right)}{a^{\frac{n+1}{m}}}$ relation in deducing this above equation. Now if we put $\bar{n}=\frac{1}{e^{\alpha k-1}}$, where $\alpha=\frac{\hbar c}{kT}$ and $\left(1+\frac{1}{\bar{n}}\right)^{-n}=\frac{1}{e^{n\alpha k}}$, using, this relation $e^{-n\alpha k}\ket{n}\bra{n}=e^{-\hat{a^{\dagger}}a\:\alpha\:k}\ket{n}\bra{n}$. Since $\displaystyle\sum_{n=0}^{\infty}\ket{n}\bra{n}=\hat{\mathds{1}}$.

So the density operator corresponding to the thermal states is

\begin{equation}
	\hat{\rho}=\frac{e^{-\hat{a^{\dagger}}a\:\alpha\:k}}{\left(\frac{e^{\alpha\:k}}{e^{\alpha\:k-1}}\right)}
\end{equation}

\section{An alternate route to our journey}

In this paper Prof. Sudarshan had given a clue that the probability distribution function can be found out by the Fourier transform of the characteristic function. Here we choose the characteristic function as  $e^{i\alpha^{*}\hat{a}^{\dagger}}e^{i\alpha\hat{a}}$. Now,

\begin{equation}
\begin{aligned}
Tr&\left[\hat{\rho}e^{i\alpha^{*}\hat{a}^{\dagger}}e^{i\alpha\hat{a}}\right] \\
&=Tr\bigg(\left[\int d^2z\ \varphi(z)\ket{z}\bra{z}\right]e^{i\alpha^{*}\hat{a}^{\dagger}}e^{i\alpha\hat{a}}\bigg) \\
&=\int d^2z \ \varphi(z)\bra{z}e^{i\alpha^{*}\hat{a}^{\dagger}}e^{i\alpha\hat{a}}\ket{z} \\
&=\int d^2z \ \varphi(z) \ e^{i\alpha^{*}z^{*}}e^{i\alpha z}
\end{aligned}
\end{equation}

If $e^{i\alpha^{*}z^{*}}e^{i\alpha z}$ is the kernel, then it represents the fourier transform of $\varphi(z)$. Now taking inverse Fourier transformation of the function $Tr\left[\hat{\rho}e^{i\alpha^{*}\hat{a}^{\dagger}}e^{i\alpha\hat{a}}\right]$ for the given density operator, we will obtain the distribution function $\varphi(z)$. Thus we get,

	\begin{equation}
		\begin{aligned}
			\varphi(z)&=\frac{1}{\pi^2}\int d^2\alpha \ tr\left(\hat{\rho} \ e^{i\alpha^{*}\hat{a}^{\dagger}}e^{i\alpha\hat{a}}\right)e^{-i\alpha^{*}z^{*}}e^{-i\alpha z} \\
			&=\frac{1}{\pi^2}\int d^2\alpha \ \sum_{n=0}^{\infty}\sum_{m=0}^{\infty}\rho_{n,m}\bra{m}e^{i\alpha^{*}\hat{a}^{\dagger}}e^{i\alpha\hat{a}}\ket{n}e^{-i\alpha^{*}z^{*}}e^{-i\alpha z} \\
			&=\frac{1}{\pi^2}\int d^2\alpha \ \sum_{n=0}^{\infty}\sum_{n^{\prime}=0}^{n}\sum_{m=0}^{\infty}\sum_{m^{\prime}=0}^{m}\rho_{n,m}\frac{\left(i\alpha^{*}\right)^{m^{\prime}}}{m^{\prime}!}\sqrt{\frac{m!}{\left(m-m^{\prime}\right)!}}\frac{\left(i\alpha\right)^{n^{\prime}}}{n^{\prime}!}\sqrt{\frac{n!}{\left(n-n^{\prime}\right)!}} \ \delta_{n-n^{\prime},m-m^{\prime}}e^{-i\alpha^{*}z^{*}}e^{-i\alpha z}
		\end{aligned}
	\end{equation}

Now readjusting the index as $n-n^{\prime}=1$ and $m-m^{\prime}=l^{\prime}$ in $\delta$- function in the above expression we get

	\begin{equation}
		\varphi(z)=\frac{1}{\pi^2}\int d^2\alpha\left[\sum_{n=0}^{\infty}\sum_{m=0}^{\infty}\sum_{K=0}^{\infty}\rho_{n+K,m+K}\frac{\sqrt{\left(n+K\right)!}\sqrt{\left(m+K\right)!}}{K!}\frac{\left(i\alpha^{*}\right)^m}{m!}\frac{\left(i\alpha\right)^n}{n!}\right]e^{-i\alpha^{*}z^{*}}e^{-i\alpha z}\label{(11)}
		\end{equation}

$\varphi(z)$ is a real function. Each term in the series is highly singular function.

\subsection{Application}
Given $\hat{\rho}=\ket{l}\bra{l}$, we can find $\varphi(z)$

We have to find first the matrix element$\hat{\rho}_{n+K,m+K}$. Here

\begin{equation}
	\begin{aligned}
		&\hat{\rho}_{n+K,m+K}=\bra{n+K}\hat{\rho}\ket{m+K} \\
		&or, \ \hat{\rho}_{n+K,m+K}=\bra{n+K}\ket{l}\bra{l}\ket{m+K} \\
		&or, \ \hat{\rho}_{n+K,m+K}=\delta_{n+K,l} \ \delta_{m+K,l} \\
	\end{aligned}
\end{equation}

Above matrix element gives non-zero vale only when $n+K=l$ and $m+K=l$ and therefore $n=m$. Hence from eq. (\ref{(11)}).

	\begin{equation}
		\varphi(z)=\frac{1}{\pi^2}\int d^2\alpha\left[\sum_{n=0}^{\infty}\sum_{m=0}^{\infty}\sum_{K=0}^{\infty}\frac{\sqrt{l!}\sqrt{l!}}{K!}\frac{\left(i\alpha^{*}\right)^m}{m!}\frac{\left(i\alpha\right)^n}{n!}\right]e^{-i\alpha^{*}z^{*}}e^{-i\alpha z}
	\end{equation}
	
Now, change the summation index  to $l-K=k$, we get

	\begin{equation}
		\varphi(z)=\frac{1}{\pi^2}\int d^2\alpha\left[\sum_{k=0}^{l}\frac{\left(-1\right)^k|\alpha|^{2k}}{k!}\cdot\frac{l!}{k!\left(l-k\right)!}\right]e^{-i\alpha^{*}z^{*}}e^{-i\alpha z}
	\end{equation}

Now, \ if $|\alpha|\rightarrow\infty$, this series will be divergent. Therefore, this Fourier transform does not exist in ordinary sense. It would appear that we cannot represent a Fock state using only diagonal expansion of coherent states. However, using $2^{nd}$ order complex Fourier transform we get,

\begin{equation*}
	\delta^{\left(2\right)}(z)=\frac{1}{\pi^2}\int d^2\alpha \  e^{-i\alpha^{*}z^{*}}e^{-i\alpha z}
\end{equation*}
Thus,
\begin{equation}
	\varphi(z)=\sum_{k=0}^{l}\frac{l!}{k!\left(l-k\right)!k!}\cdot\frac{\partial^{2k}}{\partial z^{k}\partial z^{*k}}\left(\delta^{\left(2\right)}(z)\right)
\end{equation}

This is an alternate representation of $\varphi(z)$. Let us take an example, say $l=1$, then $\hat{\rho}=\ket{1}\bra{1}$. Therefore,
\begin{equation}
	\begin{aligned}
		&\varphi(z)=\sum_{k=0}^{1}\frac{1!}{k!\left(1-k\right)!k!}\cdot\frac{\partial^{2k}}{\partial z^{k}\partial z^{*k}}\left(\delta^{\left(2\right)}(z)\right) \\
		&or, \ \varphi(z)=\delta^{\left(2\right)}(z)+\frac{\partial^{2}}{\partial z\partial z^{*}}\left(\delta^{\left(2\right)}(z)\right)
	\end{aligned}
\end{equation}

which is our required density function.

\section*{Conclusion}

In reading Prof. Sudarshan's paper it is observed that $\varphi(z)$, the quasi-probability in the diagonal representation, is real in view of the hermiticity of $\hat{\rho}$, but not necessarily positive definite. It is seen that, if we calculate the expectation values of an operator with normally ordered form, the results go hand in hand with the diagonal representation, so that the quantum mechanical expression for all correlation function will have the same form as their classical counterparts. Quasi-probability $\varphi(z)$ now playing the role played in the classical case by the true probability density of the classical ensemble. This is the essence of optical Equivalence Theorem.

All non-classicality if any, of a given state $\hat{\rho}$ are fully captured in the departure of the corresponding $\varphi(z)$ from being a genuine classical probability. This departure continues to remain the very classical definition of non-classicality. It is also clear that how the density operator can be determined from the measured correlation function, if all the correlation functions are known.

\section*{Acknowledgment}

We do hereby express our profound sense of gratitude and respect to Dr. Sobhan Sounda for spent his valuable time in helping us to finish this work. Without his proper suggestion and encouragement it was difficult for us to finish it.


\begin{thebibliography}{33}
	\expandafter\ifx\csname natexlab\endcsname\relax\def\natexlab#1{#1}\fi
	\expandafter\ifx\csname url\endcsname\relax
	\def\url#1{{\tt #1}}\fi
	\expandafter\ifx\csname urlprefix\endcsname\relax\def\urlprefix{URL }\fi
	\bibitem{E1}
	E. C. G. Sudarshan, \newblock {Phys. Rev. Lett. 10, 277 (1963)}. \label{1}
	\bibitem{E2}
	J. R. Klauder, \newblock {Ann.phys. (NY) 11, 123 (1960)}.
	\bibitem{E3}
	V. Bargmann, \newblock {Commun. Pure Appl. Math. 14, 187 (1961)}.
	\bibitem{E4}
	R. J. Glauber, \newblock {Phys. Rev. Lett. 10, 84 (1963)}. \label{2}
	\bibitem{E5}
	R. J. Glauber, \newblock {Phys. Rev. Lett. 130, 2529 (1963)}.
	\bibitem{E6}
	R. J. Glauber, \newblock {Phys. Rev. Lett. 131, 2766 (1963)}.
	\bibitem{E7}
	L. Mandel and E. Wolf, \newblock {Optical Coherence and Quantum Optics; Cambridge University Press}.
	\bibitem{E8}
	Claude Cohen-Tannoudji, Bernard Diu and Frank Laloe, \newblock {Quantum Mechanics, Volume 1}.
	\bibitem{E9}
	Walter Greiner, \newblock {Quantum Mechanics - An Introduction}.
	\bibitem{E10}
	Howard Carmichael, \newblock {Statistical Methods in Quantum Optics 1, Springer}.
	
\end{thebibliography}
\end{document}